\documentclass[a4paper]{aa}

\usepackage{graphicx,float}
\usepackage{latexsym,amsmath,amssymb}
\usepackage{natbib}
\bibpunct{(}{)}{;}{a}{}{,}

\def \arvind 	{A.~N. Parmar}
\def \david 	{D. Lumb}
\def \laurence 	{L. Boirin}
\def \mauro 	{M. Orlandini}
\def \norbert 	{N. Schartel}
\def \tim 	{T. Oosterbroek}

\def \estec {Astrophysics Mission Division, Research and Scientific
		Support Department of ESA, ESTEC, Postbus 299, NL-2200
		AG Noordwijk, The Netherlands}

\def \cesr	{Centre d'Etude Spatiale des Rayonnements, CNRS/UPS, 
		9 Av. du Colonel Roche, 31028 Toulouse Cedex 4, France}

\def \itesre	{Istituto Tecnologie e Studio Radiazioni Extraterrestri, 
		CNR, Via Gobetti 101, 40129 Bologna, Italy}

\def \vilspa 	{XMM-Newton Science Operation Center, ESA, Vilspa,
		Apartado 50727, E-28080 Madrid, Spain}

\def\persec{\hbox{s$^{-1}$}}

\def \rsun {\ifmmode$R$_{\odot}\else R$_{\odot}$\fi}
\def \nh {N${\rm _H}$}
\def \hcm {\hbox {\ifmmode $ atoms cm$^{-2}\else atoms cm$^{-2}$\fi}}

\def\approxgt{\mathrel{\hbox{\rlap{\lower.55ex \hbox {$\sim$}}
        \kern-.3em \raise.4ex \hbox{$>$}}}}
\def\approxlt{\mathrel{\hbox{\rlap{\lower.55ex \hbox {$\sim$}}
        \kern-.3em \raise.4ex \hbox{$<$}}}}

\newcommand {\sax} {BeppoSAX}

\def\arcmin{\hbox{$^\prime$}}
\def\arcsec{\hbox{$^{\prime\prime}$}}
\newcommand {\ergs} {erg~s$^{-1}$}
\newcommand {\ergcms} {erg cm$^{-2}$ s$^{-1}$}
\newcommand {\chisq} {$\chi ^{2}$}
\newcommand {\rchisq} {$\chi_{\nu} ^{2}$}

\newcommand {\phind} {$\alpha$} \newcommand {\prej} {P$_{\rm rej}$}
\newcommand {\ttnh} {$\times~$10$^{22}$~atom~cm$^{-2}$} 
\newcommand {\tunnh} {$\times~$10$^{21}$~atom~cm$^{-2}$} 
\newcommand {\ttroisnh}{$\times~$10$^{23}$~atom~cm$^{-2}$}

\newcommand {\dkpc} {d$_{\rm kpc}$}
\newcommand {\nel} {n${\rm _e}$}	
\newcommand {\mdot} {$\rm \dot{M}$}
\newcommand {\msunyear} {\rm M$_\odot$~yr$^{-1}$}

\def \src {XTE\,J0421+56}
\def \cicam {CI\,Cam}

\begin{document}

\title{Strongly absorbed quiescent X-ray emission from the X-ray
transient \src\ (\cicam) observed with XMM-Newton}

\author{\laurence\inst{1, 2}
	\and \arvind\inst{1}
 	\and \tim\inst{1}
	\and \david\inst{1}
	\and \mauro\inst{3}
	\and \norbert\inst{4}
}

\offprints{L. Boirin, \email{lboirin@rssd.esa.int}}

\institute{
	\estec
\and
	\cesr
\and
	\itesre
\and
	\vilspa
}

\date{Received 24 April 2002 / Accepted 24 July 2002}

\authorrunning{L. Boirin et al.}

\titlerunning{XMM-Newton observations of \src}

\abstract{We have observed the X-ray transient \src\ in quiescence
with XMM-Newton. The observed spectrum is highly unusual being
dominated by an emission feature at $\sim$6.5~keV.  The spectrum can
be fit using a partially covered power-law and Gaussian line model, in
which the emission is almost completely covered (covering fraction of
$ 0.98 _{-0.06}^{+0.02}$) by neutral material and is strongly absorbed
with an \nh\ of ($ 5 _{-2}^{+3}$)~\ttroisnh.  This absorption is
local and not interstellar.  The Gaussian has a centroid energy of
$6.4 \pm 0.1$~keV, a width $ \sigma <0.28$~keV and an equivalent width
of $ 940 ^{+650}_{-460}$~eV.  It can be interpreted as fluorescent
emission line from iron.  Using this model and assuming \src\ is at a
distance of 5~kpc, its 0.5--10~keV luminosity is $ 3.5 \times
10^{33}$~\ergs. The Optical Monitor onboard XMM-Newton indicates a V
magnitude of $11.86~\pm~0.03$. The spectra of X-ray transients in
quiescence are normally modeled using advection dominated accretion
flows, power-laws, or by the thermal emission from a neutron star
surface.  The strongly locally absorbed X-ray emission from
\src\ is therefore highly unusual and could result from the compact
object being embedded within a dense circumstellar wind emitted from
the supergiant B[e] companion star. The uncovered and unabsorbed
component observed below 5~keV could be due either to X-ray emission
from the supergiant B[e] star itself, or to the scattering of
high-energy X-ray photons in a wind or ionized corona, such as
observed in some low-mass X-ray binary systems.  \keywords{Accretion,
accretion disks -- Stars: individual: \src\ -- X-rays: general} }
\maketitle


\section{Introduction}
\label{sec:intro}

 \src\ was discovered by the All-Sky Monitor onboard RXTE as a soft
X-ray transient during an outburst in 1998 March 31
\citep{0421:smith98iau}. This outburst was observed by CGRO
\citep{0421:paciesas98iau}, RXTE
\citep{0421:revnivtsev99al,0421:belloni99apj}, ASCA
\citep{0421:ueda98apjl} and \sax\
\citep{0421:frontera98aa,0421:orr98aa}. The source brightened rapidly,
reaching an intensity of $\sim$2~Crab after a few hours, then quickly
decayed with an initial {\it e}-folding time of only 0.6~days before
reaching quiescence in less than 2 weeks. This was the fastest rise
and decay of any outburst from a soft X-ray transient \citep[see
e.g.,][]{chen97apj}. The outburst X-ray spectra from \src\ are complex
and can not be fit by any of the models usually applied to soft X-ray
transients.  The ASCA outburst spectrum was fitted with a two
temperature optically thin thermal model and an additional broad Fe-K
emission line at 6.4~keV.  Both \sax\ outburst spectra were described
using a two temperature bremsstrahlung model and narrow emission line
features identified with O, Ne/Fe-L, Si, S, Ca and Fe-K
\citep{0421:orr98aa}.  Emission lines at energies of $\sim$6.5~keV and
$\sim$8~keV are detected in the RXTE outburst spectra.





Optical and radio observations allowed \src\ to be rapidly identified
with \cicam, also known as MWC~84
\citep{0421:wagner98iau,0421:hjellming98iau,0421:robinson98iau}.  It
is a frequently observed source in ultra-violet (UV), optical and
infra-red (IR) wavelengths.  Its V magnitude over long term
observations shows a $\sim$0.4~magnitude amplitude variability and has
a mean value of 11.6, both before (1989--1994) and after (1998--1999)
the 1998 X-ray outburst \citep{bergner95aa,0421:clark00aa}.  \cicam\
is a supergiant B[e] star \citep{clark99aa,0421:robinson02apj}, or a
sgB[e] star, following the notation of \citet{lamers98aa}, i.e. a
supergiant showing the B[e] phenomenon.  The B[e] phenomenon concerns
many objects of different masses and evolutionary phases \citep[see
e.g.,][]{lamers98aa}. One of the common properties of stars exhibiting
the B[e] phenomenon is the presence of forbidden emission lines in
their optical spectra (the notation ``[e]'' refers to the one used for
forbidden lines). Another common property is to show a strong IR
excess attributed to hot circumstellar dust. In these respects, stars
with the B[e] phenomenon clearly differ from the ordinary Be stars
which are rapidly rotating stars near the main sequence losing mass in
an equatorial wind. In practice, the spectroscopic and photometric
properties of stars with the B[e] phenomenon are also easily
distinguished from those of ordinary Be stars. \cicam/\src\ is the
first high-mass X-ray binary (HMXB) with a sgB[e] mass donor
companion. Another source suspected to be a HMXB with a mass donor
showing the B[e] phenomenon is the optical/X-ray source
HD~34921/1H~0521+37 \citep{clark99aa}.  Adopting the
classification criteria and notation of \citet{lamers98aa},
\citet{clark99aa} identify the companion star in this system as an
``unclB[e] star'' (unclassified B[e] star).


    Optical high-dispersion spectroscopy of \cicam\ led
 \citet{0421:robinson02apj} to the conclusion that the sgB[e] star
 emits a two component wind.  One component is a hot, high-velocity
 wind. The other component is a cool, low-velocity and very dense
 (electron number density log~\nel~$>~9.5$) wind. The wind is roughly
 spherical and continuously replenished. The mass-loss rate due to the
 wind is very high: \mdot~$> 10^{-6}$~\msunyear. This wind fills the
 space around the sgB[e] star and, from the size of the IR-emitting
 dust shell, extends to a radius between 13 and 50~AU.  Thus, the
 circumstellar material around \cicam\ is much denser, far more
 extended, and much less confined to the equatorial plane than the
 circumstellar material around a Be star.  The passage of the compact
 X-ray source through such a complex and dense environment is likely
 to strongly affect the X-ray properties of the source.

 Within this picture, \citet{0421:robinson02apj} suggest that the 1998
outburst was caused by the same disk instability mechanism responsible
for the outbursts in X-ray novae, i.e. by an instability in the
accretion disk around the compact object \citep[see
e.g.,][]{lasota01nar}. It would thus differ from the outbursts
observed in ordinary Be HMXB that recur at multiples of the orbital
period, when the compact object comes close to the Be star at
periastron and plunges into its equatorial wind.



The distance to \src\ is uncertain.  Based on optical spectroscopic
properties, on radial velocity measurements of \cicam\ and on
considerations about the structure of the Galaxy,
\citet[Sect.~2.3]{0421:robinson02apj} estimate that the distance to
the source is much larger than the $\lesssim$2~kpc previously
considered.  \citet{0421:robinson02apj} use a distance of 5~kpc and
note that it is likely to be a lower limit to the true distance which
could be up to 10~kpc. In this paper, we assume a distance of 5~kpc.
This distance makes \src\ among the most luminous transients. The
2--25~keV luminosity at the peak of the outburst was $3.0 \times
10^{38}$~\ergs, assuming the revised distance of 5~kpc
\citep{0421:orlandini00aa,0421:robinson02apj}.

The unusual nature of \cicam\ makes the interstellar absorption
towards the star difficult to estimate.  From an UV spectrogram,
\citet{0421:robinson02apj} derive a differential extinction E(B--V) of
$0.85 \pm 0.05$, but do not attempt to separate circumstellar from
interstellar extinction.  From an analysis of diffuse interstellar
bands in the optical spectrum of \cicam, \citet{0421:clark00aa} derive
an interstellar E(B--V) of $0.65 \pm 0.20$ and an ${\rm A_v}$ of $2.0
\pm 0.6$, which implies an interstellar X-ray absorption, \nh, of
$(3.6 \pm 1.1)$~\tunnh\ \citep[Sect. 3]{0421:parmar00aa}.  Extinction
at soft X-ray wavelengths yielded an \nh\ of $(3.76 \pm 0.36)$~\ttnh\
near the peak of the outburst, and the \nh\ decreased to
$\sim$2.2~\tunnh\ as \src\ approached quiescence
\citep{0421:belloni99apj}. This rapid change in the X-ray extinction,
as well as the change in the IR flux after the outburst
\citep{0421:clark00aa}, indicate that much of the extinction to
\cicam\ is local, not interstellar \citep{0421:robinson02apj}.

No bursts, pulsations or quasi-periodic oscillations have been
  detected from \src\ \citep{0421:belloni99apj}.  The large ratio of
  peak to quiescent luminosity is taken as evidence for the compact
  object being a black hole \citep{0421:robinson02apj}. Furthermore,
  \cicam\ has been detected as a relatively bright radio source
  \citep{0421:hjellming98iau} which is more typical of black hole
  candidates than neutron star systems
  \citep[see][Sect. 5]{0421:belloni99apj}.

\src\ was observed in quiescence by \sax\ on 1998 September 03, 1999
September 23 and 2000 February 20
\citep{0421:orlandini00aa,0421:parmar00aa}. In 1998, the source was
soft (power-law photon index, \phind, of $4.0 ^{+1.9} _{-0.9}$) with a
low \nh\ of $(1 ^{+5} _{-1}$)~\tunnh. In 1999, the source had
hardened (\phind~$= 1.86 ^{+0.27} _{-0.32}$) and brightened and became
strongly absorbed with an \nh\ of $(4.0~\pm~0.8)$~\ttroisnh.  There is
evidence for a narrow emission line in both spectra at $\sim$7~keV. In
2000, the source was not detected.  At 5~kpc, the 1--10~keV
luminosities were $1.4 \times 10^{33}$, $2.3 \times 10^{34}$, and $<$$2.5 \times 10^{33}$~\ergs, in 1998, 1999, and 2000, respectively.
These results are summarized in Table \ref{tab:sumxobs}.

\begin{table}
\caption{Summary of quiescent X-ray observations of \src. The
columns indicate respectively the observatory (SAX for \sax, XMM for
XMM-Newton), the year of observation, the Hydrogen column density
derived, the 1--10~keV luminosity at 5~kpc, and the references ([1]
for Orlandini et al. (2000), [2] for Parmar et al. (2000), [3] for
this work). }
\begin{tabular}{lllll}
\hline
\hline
Obs. & Year & \nh &  L$_{\rm 1-10\; keV}$  & Ref.\\
 & &  (atom cm$^{-2}$) &  (\ergs)  & \\
\hline
SAX & 1998 & $(1 ^{+5} _{-1}) \times  10^{21}$ & $1.4 \times 10^{33}$ & [1], [2]  \\
SAX & 1999 & $(4.0~\pm~0.8) \times  10^{23} $& $2.3 \times 10^{34}$ & [2] \\
SAX & 2000 & & $<$$2.5 \times 10^{33}$& [2] \\
XMM & 2001 & $(5^{+3}_{-2}) \times  10^{23}$& $3.3 \times 10^{33}$ & [3]\\
\hline
\hline
\end{tabular}
\label{tab:sumxobs}
\end{table}

Here, we report on the XMM-Newton observation of \src\ in quiescence
performed on 2001 August 19. We present and discuss the nature of the
X-ray spectrum and derive the V magnitude of the source using the
Optical Monitor.


\section{Observations and data analysis}

The XMM-Newton Observatory \citep{jansen01aa} includes three
1500~cm$^2$ X-ray telescopes each with an European Photon Imaging
Camera (EPIC, 0.1--15~keV) at the focus.  Two of the EPIC imaging
spectrometers use MOS CCDs \citep{turner01aa} and one uses pn
CCDs \citep{struder01aa}. Reflection Grating Spectrometers \citep[RGS,
0.35--2.5~keV,][]{denherder01aa} are located behind two of the
telescopes. In addition, a coaligned optical/UV Monitor \citep[OM,
160--600~nm,][]{mason01aa} is included as part of the payload.

The region of sky containing \src\ was observed by XMM-Newton on 2001
August 19 between 07:05 and 16:16 UTC.

\subsection{X-ray observations}

\begin{figure}[!t]

\centerline{\includegraphics[width=0.5\textwidth]{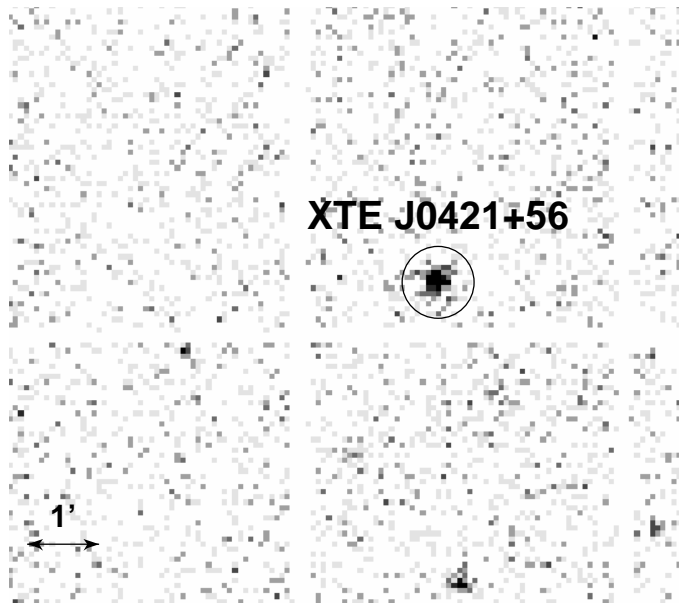}}

\caption{An EPIC pn image of the region of sky containing \src. The
radius of the circle showing the source position is 30\arcsec. North
is up and East is to the right.}

\label{fig:image}

\end{figure}

The X-ray cameras were operating in their Prime Full Window mode with
Medium thickness filters.  We used the X-ray data products generated
by the Pipeline Processing Subsystem  in September 2001. We
further filtered these products using the Science Analysis Software
(SAS) version 5.1.0\footnote{ After the public release of SAS
version 5.3.0, we used this latest version and re-extracted the pn
spectrum. The obtained spectrum is identical to the one obtained using
SAS version 5.1.0.} and especially the tasks {\it evselect} and {\it
xmmselect}. Electronic noise and hot or flickering pixels were
rejected.  The observation is contaminated by high backgrounds
intervals due to solar activity. We excluded these times by selecting
intervals where the overall $>1$~keV pn count rate was $<$$15~\rm
s^{-1}$.  Source counts were extracted in a 30\arcsec\ radius
circular region centered on \src. The source is close to a chip
border in the pn detector. Background counts were obtained from a
circular region of 90\arcsec\ radius offset from the source position.
For the pn, single and double pixel events were selected
(patterns 0 to 4). For the MOS, events corresponding to patterns from
0 to 12 were selected. The exposure after applying these filtering
criteria is 12~ks. This is substantially below the expected
exposure due to the removal of intervals of high solar
activity.  Fig~\ref{fig:image} shows the EPIC pn image of the
region of sky containing \src. The source,  although faint, is
clearly detected.

We next extracted spectra.  For the pn, subtracting the source and
background count rates gives a net count rate of
$0.024~\pm~0.002$~\persec\ in the 0.1--12~keV energy range, where the
source is detected. This corresponds to $\sim$280 net counts detected.
We used the standard pn response matrix file
epn\_ff20\_sY9\_medium.rmf.  Approximately 60 net counts are detected
in each MOS camera. We used the response matrix files
m1\_medv9q19t5r5\_all\_15.rsp and m2\_medv9q19t5r5\_all\_15.rsp for
MOS~1 and MOS~2, respectively.  In order to ensure applicability of
the \chisq\ statistic with so few counts, we rebinned the MOS and pn
spectra such that at least 25 net counts per bin were present. The
resulting MOS spectra have 2 bins and the pn spectum has 11 bins,
allowing only simple models to be tested. \src\ is not detected in the
RGS.  Modeling of the spectra was carried out using XSPEC version
11.1.0. All spectral uncertainties are given at 90\% confidence and
upper limits at 95\% confidence.

\subsection{Optical observations}

The OM was operating in its default Imaging Mode.  Five consecutive
exposures of 1000~s were obtained through the V filter (510-580~nm),
followed by five consecutive exposures of 4540~s through the UVW1
filter (245-320~nm). We have analyzed 2\arcmin$\times$2\arcmin\ high
resolution images centered on \src.

Instrumental magnitudes of the source were extracted for each image
using DAOPHOT.  We computed the average instrumental magnitudes for
both filters and used these values to correct the average V
instrumental magnitude into a standard V magnitude of the Johnson's
UBV system. We used the color transformation relation shown e.g., in
\citet{antokhin02conf} and the coefficients from the Current
Calibration File OM\_COLORTRANS\_0006.CCF.

\section{Results}

\subsection{The X-ray spectrum}
\label{sec:spectrum}

\begin{figure}[!t]

\centerline{\rotatebox{-90}{\includegraphics[height=0.32\textheight]{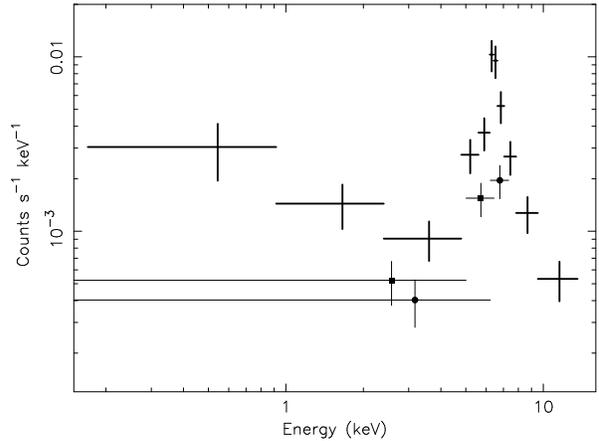}}}

\caption{The MOS~1 (squares), MOS~2 (circles) and pn (thick
crosses) count spectra of \src\ in quiescence.}

\label{fig:spectrumcomb}
\end{figure}

The combined MOS~1, MOS~2 and pn spectrum of \src\ is shown in
Fig.~\ref{fig:spectrumcomb}.  Its shape is highly unusual. It is
clearly dominated by an emission feature peaking at
$\sim$6.5~keV\footnote{As \src\ is located near a chip border in
the pn detector, which could produce artefacts in the spectrum
\citep[see e.g.,][]{borozdin02apjl}, we checked that the 6.5~keV
feature was not due to the closeness of the edge. We extracted a pn
spectrum from a 30\arcsec\ radius circle located at the same distance
from the edge as the source. This spectrum does not show a feature
around 6.5~keV, indicating that this feature is not an artefact caused
by the closeness to a chip border.}.

We first tried to fit the spectrum with simple descriptive models.  A
 model consisting of a power-law with a photon index, \phind\,
 $\sim$~-0.9, to account for the low-energy feature of the spectrum
 ($\lesssim$5~keV) and a Gaussian to account for the high-energy
 feature (above $\gtrsim$5~keV) does not provide an acceptable fit to
 the data giving a reduced \chisq\ (\rchisq) of 2.07 for 10
 degrees of freedom (d.o.f.).  A similarly poor result is
 obtained when including photo-electric absorption ({\it phabs} within
 XSPEC) to this model in order to account for an absorption of the
 X-rays.  Both a broad and a narrow component seem to be needed to
 account for the high-energy feature. A relatively good (\rchisq\
 of 1.56 for 8 d.o.f.) description of the spectrum can be obtained
 using a model consisting of a power-law (\phind~$\sim$~1.9) to
 account for the low-energy feature plus another power-law
 (\phind~$\sim$~-1.3) to account for the broad high-energy feature and
 a Gaussian at $\sim$6.4~keV to account for the narrow high-energy
 feature.

We then adopted a different fitting approach by modeling the broad
high-energy feature as a highly absorbed continuum component.  We
tried a model consisting of a power-law for the low-energy feature,
plus a power-law and a Gaussian both modified by absorption from
neutral material for the high-energy feature (model {\it
powerlaw+phabs(powerlaw+gaussian)} within XSPEC). This model fits the
spectrum relatively well, with a \rchisq\ of  1.04 for 7 d.o.f..
The best-fit parameters for this model are given in Table
\ref{tab:resultscomb}.  We note that, without the Gaussian in the
model, the fit is poor, with a \rchisq\ of 2.19 for 10 d.o.f.. An
F-test indicates that the probability, \prej, of rejecting the
hypothesis that the fit is better including the Gaussian is 
4.3\%. This indicates that the Gaussian is significant at 
95.7\% confidence.  In order to account for potential absorption of
the first power-law emission by the line of sight interstellar medium,
we included an additional photo-electric absorption component ({\it
phabs} within XSPEC) to the previous model. The resulting model ({\it
phabs(powerlaw+phabs(powerlaw+gaussian))} fits the spectrum relatively
well, with a \rchisq\ of 1.22 for 6 d.o.f., although an F-test
indicates that the addition of the absorption component does not
improve the fit (\prej\ of 99.9\%). The best-fit parameters
obtained using this model are also given in Table
\ref{tab:resultscomb}. The \nh\ of the additional absorption component
is $<$1.8~\tunnh. The other parameter values are consistent with those
obtained previously without the additional {\it phabs} component.

 We then tried to fit the spectrum using a partially covered
power-law and Gaussian model, the {\it pcfabs(powerlaw+gaussian)}
model within XSPEC. In this model, some absorbing material covers a
fraction (from 0 to 1) of the power-law  and Gaussian
emission. The absorption is by neutral material with solar
abundances and uses cross-sections from \citet{balucinska92apj}. 
This model fits the spectrum very well with a \rchisq\ of 1.04 for 8
d.o.f. The best-fit parameters are given in
Table~\ref{tab:resultscomb}.  Without the Gaussian in the model, the
fit is poor, with a \rchisq\ of 2.47 for 11 d.o.f. An F-test indicates
that the probability, \prej, of rejecting the hypothesis that the fit
is better including the Gaussian is 1.8\%. This indicates that the
Gaussian is significant at 98.2\% confidence.  In order to account for
potential interstellar absorption of the uncovered component of the
emission, we included an additional photo-electric absorption
component to the previous model. The resulting model ({\it
phabs(pcfabs(powerlaw+gaussian))} fits the spectrum relatively well,
with a \rchisq\ of 1.18 for 7 d.o.f., although an F-test indicates
that the addition of the absorption component does not improve the
fit, with \prej\ of 99.7\%. The best-fit parameters obtained using
this model are given in Table~\ref{tab:resultscomb}. The \nh\ of the
additional absorption component is $<$2.0~\tunnh. The other parameter
values are consistent with those obtained previously, without the
additional {\it phabs} component.

\begin{table*}

\caption{Fit results of the combined MOS~1, MOS~2 and pn
spectrum.  The names of the models follow the XSPEC syntax. {\it
phabs} is photo-electric absorption. {\it pcfabs} is partial covering
fraction absorption. {\it pl} designates {\it powerlaw} and is a
power-law with a photon index, \phind. {\it gauss} designates {\it
Gaussian} and is a simple Gaussian line profile.  \rchisq\ is the
reduced \chisq.  \nh\ is the Hydrogen column density. CvrFract is the
covering fraction (from 0 to 1) in the {\it pcfabs} model.}

\begin{center}
\begin{tabular}{lll}
\hline
\hline
\multicolumn{3}{c}{MOS 1 + MOS 2 + pn spectrum}\\
\hline
\hline
Model & {\it pl+phabs(pl+gauss)} & {\it phabs(pl+phabs(pl+gauss))}\\
\rchisq\ (d.o.f.) &  1.04 (7) & 1.22 (6)\\
\hline
\nh\ (atom cm$^{-2}$) & \dots & $<$$1.8 \times 10^{21}$ \\
\phind\ & $1.2^{+0.6}_{-0.7}$ & $1.2^{+1.1}_{-0.8}$\\
\nh\ (atom cm$^{-2}$) & ($5 \pm 2$)~$\times 10^{23}$ & ($5^{+3}_{-2}$)~$\times 10^{23}$ \\
\phind\ & $1.0 \pm 0.9$ & $1.0^{+1.2}_{-0.9}$\\
Gaussian energy  & $6.4 \pm 0.1 $~keV & $6.4 \pm 0.1$~keV\\
Gaussian width $\sigma$  & $<$$0.29$~keV & $<$$0.29$~keV\\
Gaussian equivalent width & $1010^{+720}_{-480}$ eV &  $1010^{+720}_{-480}$ eV\\
\hline
\hline
Model & {\it pcfabs(pl+gauss)} & {\it phabs(pcfabs(pl+gauss))} \\
\rchisq\ (d.o.f.) & 1.04 (8) & 1.18 (7) \\
\hline
\nh\ (atom cm$^{-2}$)  & \dots & $<$$2.0 \times 10^{21}$ \\
\nh\ (atom cm$^{-2}$) & ($5 _{-2}^{+3}$)~$\times 10^{23}$ & ($5^{+3}_{-2}$)~$\times 10^{23}$ \\
CvrFract & $0.98 _{-0.06}^{+0.02}$ &$0.98^{+0.02}_{-0.06}$\\
\phind\ & $1.2 \pm 0.7$ & $1.2^{+1.1}_{-0.7}$\\
Gaussian energy  & $6.43^{+0.10}_{-0.09} $~keV & $6.43 \pm 0.09 $~keV\\
Gaussian width $\sigma$  & $<$0.28 keV & $<$0.28 keV\\
Gaussian equivalent width & $940^{+650}_{-460}$ eV &  $940^{+650}_{-460}$ \\
\hline
\hline
\end{tabular}
\end{center}

\label{tab:resultscomb}
\end{table*}

Summarizing, we have obtained acceptable fits to the combined pn and
MOS spectra of XTE J0421+560 in quiescence using two different two
component models. One component is strongly absorbed and the other may
be unabsorbed, or only slightly absorbed (Table
\ref{tab:resultscomb}).  In both models, the shape of the high-energy
continuum is dominated by absorption resulting in a strong cutoff
above 7.1 keV due to neutral iron absorption edge. This produces a
very sharply peaked spectral feature which resembles a broad emission
line. In addition, there is some evidence for the presence of an
additional narrow emission feature at an energy of 6.4 keV. (This
feature is required at 98.2\% confidence in the case of the partially
covered power-law and Gausssian model). The two models presented in
Table 2 differ by how the low-energy component is modeled. This is
either by partial covering, in which case the low-energy component is
constrained to have the same underlying spectral shape as the absorbed
high-energy component, or by a power-law with an unconstrained
slope. For completeness, Table~\ref{tab:resultscomb} lists the results
if additional low-energy absorption is included to account for
interstellar absorption. However, the fits do not require such
absorption, and the upper-limits are consistent with that expected
from the likely interstellar values.

\begin{figure*}[!ht]
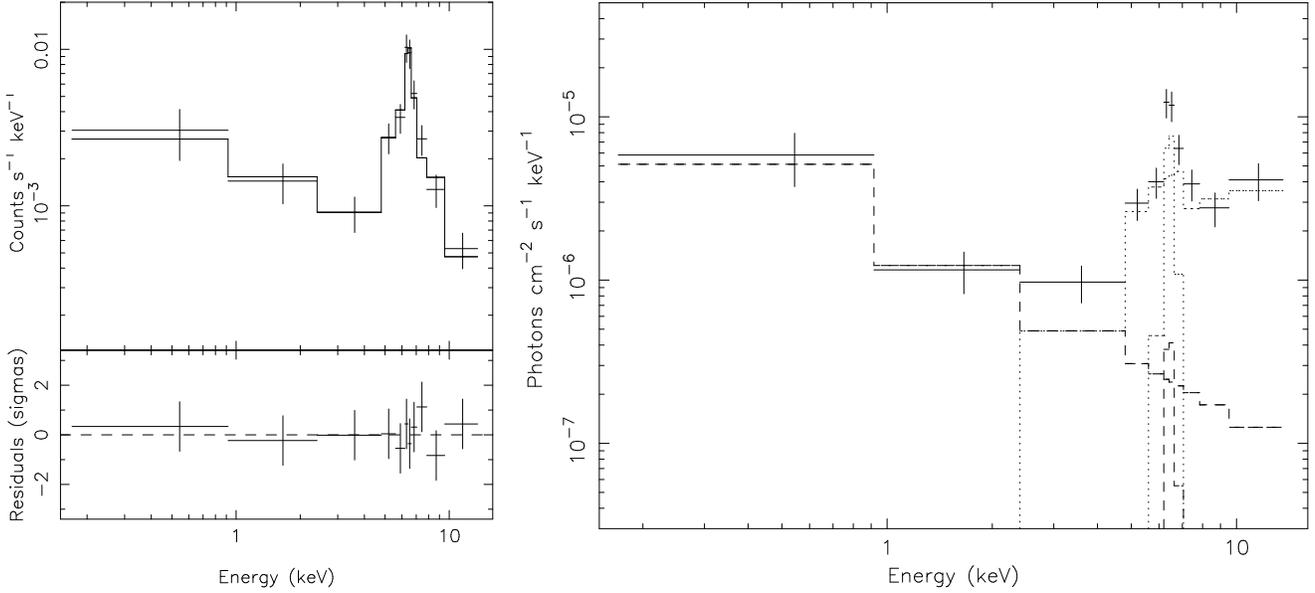


\centerline{\hfill\includegraphics[height=0.32\textheight]{H3654F3a.ps}\hfill\rotatebox[origin=rb]{-90}{\includegraphics[width=0.32\textheight]{H3654F3b.ps}}\hfill}


\caption{The spectrum of \src\ in quiescence (only the pn is
shown for clarity). The left panels show the count spectrum (data and
folded model). The solid line is the best-fit using the partially
covered power-law and Gaussian model. The lower left panel shows the
residuals from the fit in terms of  standard deviations.  The
right panel shows the unfolded spectrum. The dashed line shows the
contribution of the unabsorbed component (the low-energy component) to
the total model. The dotted line shows the contribution of the
absorbed component (the high-energy component). The total model is not
shown in this panel for clarity.}

\label{fig:spectrum}
\end{figure*}

 The partially covered power-law and Gaussian model and the
 power-law plus absorbed power-law and Gaussian model give equally
 good fits with a \rchisq\ of 1.04 for 8
 d.o.f.. Fig.~\ref{fig:spectrum} shows the former model fit to the pn
 spectrum only for clarity.  In this model , the power-law and
 Gaussian emissions are almost completely covered (with a covering
 fraction of 0.98) by neutral material and strongly absorbed with an
 \nh\ of ($5 _{-2}^{+3}$)~\ttroisnh. The equivalent width of the
 Gaussian line feature is $940 ^{+650}_{-460}$~eV.  This model is
 equivalent to the sum of an absorbed (and covered) plus an unabsorbed
 (and uncovered) component with the same spectral shape.  The
 contributions of both these components are shown separately in the
 right panel of Fig.~\ref{fig:spectrum}.  The unabsorbed omponent
 (dashed line) dominates the emission $\lesssim$5~keV. We refer to
 this component as the low-energy component.  On the contrary, the
 absorbed component (dotted line) clearly dominates the emission
 $\gtrsim$5~keV.  We refer to this component as the high-energy
 component.  Using the partially covered power-law and Gaussian
 model, we derive a 0.5--10~keV absorbed flux of $ 2.8\times
 10^{-13}$~\ergcms\ and a unabsorbed flux of $ 1.2\times
 10^{-12}$~\ergcms\ by setting \nh\ to 0. This corresponds to an
 unabsorbed luminosity of $ 1.4 \times 10^{32}$~\dkpc$^{2}$~\ergs\
 at a distance of \dkpc\ given in kpc, and to a unabsorbed luminosity
 of $ 3.5\times 10^{33}$~\ergs, at a distance of 5~kpc. The
 low-energy component contributes  9.8\% to the total absorbed
 flux in the  0.5--10~keV range.   The 1--10~keV luminosity
 is given in Table~\ref{tab:sumxobs} for comparison with previous
 \sax\ observations of \src\ in quiescence.

\subsection{The V magnitude}

The average UVW1 instrumental magnitude of \src\ during the XMM-Newton
observation is $12.48 \pm 0.03$.  The average standard V magnitude is
$11.86 \pm 0.03$.


\section{Discussion}

\subsection{The V magnitude}

Optical observations of \src\ (\cicam) exist prior to the 1998
outburst.  A V magnitude of 11.4 is attributed to \cicam\ during
observations made in the 1970's \citep{allen76aa}.

Long term observations of the source were carried out between 1989 and
1994. \cicam\ covered a range of V magnitudes between $11.44 \pm 0.03$
and $11.76 \pm 0.03$ (47 data points), thus showing a
$\sim$0.4~magnitude amplitude variability on a day-to-day timescale
\citep{bergner95aa}.  The mean V magnitude of \cicam\ during this
pre-outburst interval was $11.634 \pm 0.004$.  Performing a Fourier
analysis on these data, \citet{0421:miroshnichenko95aat} found
evidence for a 11.7~days quasi-period which he interpreted as a
possible orbital period.

\citet{0421:clark00aa} report on observations of \cicam\ performed
after the outburst, between 1998 August and 1999 March.  The V
magnitude of \cicam\ during this post-outburst interval ranges between
$11.52 \pm 0.04$ and $11.70 \pm 0.01$ (18 data points) and shows the
same $\sim$0.4~magnitude variability with the same mean value as
during the pre-outburst period.

Thus, the V magnitude of $11.86 \pm 0.03$ during the XMM-Newton
observations is outside the range of magnitudes reported previously,
indicating that the source had become fainter.  However, this
magnitude represents only one data point.  So, it is difficult to
determine if this higher value is due to variability on a day-to-day
timescale, or if the source has become fainter for an extended
interval.

\subsection{The X-ray emission}

The quiescent XMM-Newton spectrum of \src\ can be fit by a
power-law plus absorbed power-law and Gaussian model or alternatively
by a partially covered power-law and Gaussian model. In both models,
the high-energy component corresponds to a strongly absorbed continuum
(${\rm N_H}$ of $ 5 _{-2}^{+3}$~\ttroisnh).  This fitting approach
is supported by the fact that a similarly strongly absorbed quiescent
emission was also reported from the source during the 1999 \sax\
observation.  \citet{0421:parmar00aa} fit the \sax\ 2--10~keV spectrum
with an absorbed (\nh\ of $(4.0 \pm 0.8)$~\ttroisnh) power-law
(\phind~$= 1.86 ^{+0.27}_{-0.32}$) together with a Gaussian emission
feature with an energy of 7.3~$\pm$~0.2~keV and an equivalent width of
620~$\pm$~350~eV. Since an unabsorbed component is unambiguously
detected at low-energy in the XMM-Newton quiescent spectrum of
\src, we re-examined the \sax\ spectrum reported in
\citet{0421:parmar00aa} to see if there is evidence for the presence
of a similar component. The 0.2--2~keV LECS count rate of
(7.5~$\pm$~4.8)~$\times$~10$^{-4}$~\persec\ suggests that such a
component may be present. In order to investigate this further, we fit
the partially covered XMM-Newton model discussed above to the
0.5--10~keV \sax\ spectrum allowing the spectral parameters to
vary. The line energy and width which were poorly constrained were
set to the best-fit XMM-Newton values.  This gives a \rchisq\ of 1.43
for 38 d.o.f.. The uncovered component contributes (6$\pm$3)\% of the
total 1--10~keV absorbed flux, consistent with the ratio observed with
XMM-Newton.  Thus, the presence in the \sax\ 1999 spectrum of an
unabsorbed component, similar to that detected in the XMM-Newton
spectrum cannot be excluded.

When absorption is added to either the power-law plus absorbed
power-law and Gaussian model or the partially covered power-law and
Gaussian model used to fit the XMM-Newton spectrum, the \nh\ of this
component is $<$2.0~\tunnh\ (see Table~\ref{tab:resultscomb}).
This value is consistent with that obtained when \src\ approached
quiescence after the outburst \citep{0421:belloni99apj}.  These low
values of \nh\ confirm that the {\it interstellar} column towards
\src\ is not high.  On the other hand, X-ray results show that the
column density intrinsic to the system can be very high, and as
inferred from the large range of absorption obtained (from roughly 0.2
to 50~\ttnh), very variable. Thus, this confirms the picture that most
of the absorption towards \src\ is local and not interstellar.

The Gaussian feature observed at 6.4~keV can be interpreted as a
fluorescent emission from iron. Such an interpretation is consistent
with the modeling of the spectrum using partial covering since it
suggests the presence of significant cold absorbing material in the
system. The large equivalent width observed can be explained if cold
material surrounds the X-ray emitter with a large column density,
which is also consistent with our modeling and with the picture, drawn
from optical spectroscopy, of a compact object embedded in a very
dense wind \citep{0421:robinson02apj}.  However, the spectral quality
is too low to test for the presence of associated signatures of cold
material such as absorption edges. A reflection component due to the
presence of cold material could be expected as well, but such a
component usually peaks above the energy range covered here (between
$\sim$10 and 100~keV). Gaussian emission features were detected
during the 1998 and 1999 \sax\ observations of \src\ in quiescence at
energies of $7.0^{+1.6}_{-0.2}$ and 7.3~$\pm$~0.2~keV
respectively. Their different energies as compared to the 6.4~keV
feature detected with XMM-Newton may indicate a different origin, or
different physical conditions in the emission region.  Emission
features at $\sim$6.4~keV were also detected in outburst spectra from
\src\ and mostly interpreted as emission lines produced by an
optically thin plasma \citep[see
e.g.,][]{0421:ueda98apjl,0421:revnivtsev99al}. Such a mechanism can
not be excluded in the case of the XMM-Newton observation of \src,
although the geometry and emission processes involved during
quiescence are likely to be very different from those involved during
the outburst.

The partially covered power-law  and Gaussian model as well as the
powerlaw plus absorbed power-law and Gaussian model suggest the
presence of two components: an unabsorbed component mainly observed
$\lesssim$5~keV (the uncovered or low-energy component), and second, a
strongly absorbed component mainly observed $\gtrsim$5~keV (the
covered or high-energy component) and dominating the total spectrum.

We propose that the covered component results from the compact object
being embedded within the dense circumstellar wind emitted from the
sgB[e] companion star, in agreement with the picture drawn from
optical spectroscopy of the source \citep{0421:robinson02apj}.  The
large range of observed \nh\ at X-ray wavelengths could reflect the
complexity of the B[e] star environment in which the compact object is
traveling. Regions with different physical properties may be crossed,
depending e.g., on the distance of the compact object from the sgB[e]
star or from its equatorial plane. This environment may vary with time
as well. It may also be modified by the X-rays emitted in the vicinity
of the compact object, which are probably variable themselves.  

We suggest two possible origins for the low-energy component.  First,
it could be due to X-ray emission from the sgB[e] star itself.  The
X-ray emission from OB stars is intrinsically soft \citep[up to
$\sim$4~keV,][]{long80apjl}.  \citet{0421:orlandini00aa} estimate that
the X-ray luminosity of the companion star in \src\ could be
$\sim$5~$\times 10^{32}$~\ergs, while
\citet[Sect. 2.4]{0421:robinson02apj} estimate that the sgB[e] star
could emit up to $10^{34}$~\ergs\ in the 0.2--4.0~keV band.  The
0.2--4.0~keV luminosity observed from \src\ during the XMM-Newton
observation is $1.5 \times 10^{33}$~\ergs\ at 5~kpc. Thus, we cannot
exclude that the low-energy emission, or part of it, originates from
the companion star.

Another possibility is that the low-energy component is due to the
scattering of higher-energy X-ray photons  in a wind or ionized
corona such as observed in some low-mass X-ray binaries.  The flux of
the low-energy component in \src\ is about 10\% of the total
 0.5--10~keV flux.  In dipping, eclipsing or accretion disk
corona sources, the ratio observed between the flux attributed to
scattered emission and the total flux is usually $\sim$5\% \citep[see
e.g.,][]{0748:parmar86apj}. Thus, at least a part of the low-energy
emission could be due to scattering in \src.   We note however
that corona have been observed in low-mass X-ray binaries that are
much brighter than \src. So the possible scattering region in \src\
may differ in nature and formation from those observed in low-mass
X-ray binaries. The scattering region in \src\ could be linked to the
wind emitted by the B[e] companion star.  Emission from the companion
star and scattering could both play a role in the low-energy emission
observed from \src.

\src\ is the first identified member of a new class of HMXB with
sgB[e] companion. It is the only known system in which the compact
object is immersed in a dense and complex circumstellar wind.  Further
multiwavelength observations of this source are needed to explore the
geometry and the emission processes involved in this system.  Many
other stars showing the B[e] phenomenon, and especially sgB[e] stars,
could host a compact object. Due to their low X-ray luminosity and
absorbed spectra, such objects are unlikely to have been identified in
previous low-energy (0.1--2.5~keV) sky surveys such as conducted by
ROSAT, and we await future medium energy X-ray surveys to detect
further members of this class.




\begin{acknowledgements}

This work is based on observations obtained with XMM-Newton, an ESA
science mission with instruments and contributions directly funded by
ESA member states and the USA (NASA).  L.~Boirin acknowledges an ESA
Fellowship.  We thank Rudi Much, Igor Antokhin and Simon Rosen for
providing very useful help analyzing the XMM-Newton Optical Monitor
data.





\end{acknowledgements}


\bibliographystyle{aa}

\end{document}